%% file: paper.tex
\documentclass[conference]{IEEEtran}
\IEEEoverridecommandlockouts
\usepackage{cite}
\usepackage{amsmath,amssymb,amsfonts}
\usepackage{algorithmic}
\usepackage{graphicx}
\usepackage{textcomp}
\usepackage{xcolor}
\def\BibTeX{{\rm B\kern-.05em{\sc i\kern-.025em b}\kern-.08em
    T\kern-.1667em\lower.7ex\hbox{E}\kern-.125emX}}
\begin{document}

\title{Memory Is All You Need: \\
An Overview of Compute-in-Memory Architectures for Accelerating Large Language Model Inference}

\author{\IEEEauthorblockN{Christopher Wolters\textsuperscript{1,2} \qquad Xiaoxuan Yang\textsuperscript{3} \qquad
Ulf Schlichtmann\textsuperscript{1} \qquad Toyotaro Suzumura\textsuperscript{2}}
\IEEEauthorblockA{\textsuperscript{1}Technical University of Munich \qquad
\textsuperscript{2}The University of Tokyo \qquad
\textsuperscript{3}Stanford University \\ [0.15cm]
{\tt\small ch.wolters@tum.de}}}

\maketitle

\begin{abstract}
Large language models (LLMs) have recently transformed natural language processing, enabling machines to generate human-like text and engage in meaningful conversations. This development necessitates speed, efficiency, and accessibility in LLM inference as the computational and memory requirements of these systems grow exponentially. Meanwhile, advancements in computing and memory capabilities are lagging behind, exacerbated by the discontinuation of Moore's law. With LLMs exceeding the capacity of single GPUs, they require complex, expert-level configurations for parallel processing. Memory accesses become significantly more expensive than computation, posing a challenge for efficient scaling, known as the memory wall. 
Here, compute-in-memory (CIM) technologies offer a promising solution for accelerating AI inference by directly performing analog computations in memory, potentially reducing latency and power consumption. By closely integrating memory and compute elements, CIM eliminates the von Neumann bottleneck, reducing data movement and improving energy efficiency. \\
This survey paper provides an overview and analysis of transformer-based models, reviewing various CIM architectures and exploring how they can address the imminent challenges of modern AI computing systems. We discuss transformer-related operators and their hardware acceleration schemes and highlight challenges, trends, and insights in corresponding CIM designs.
\end{abstract}


\input{introduction}
\input{background}
\input{challenges}
\input{approaches}
\input{discussion}
\input{conclusion}

\newpage

\bibliographystyle{IEEEtran}
\bibliography{references}
\end{document}

%% file: introduction.tex
\section{Introduction}
\subsection{Advancements in Large Language Models}
\begin{figure}[b]
\centering
\includegraphics[width=\linewidth]{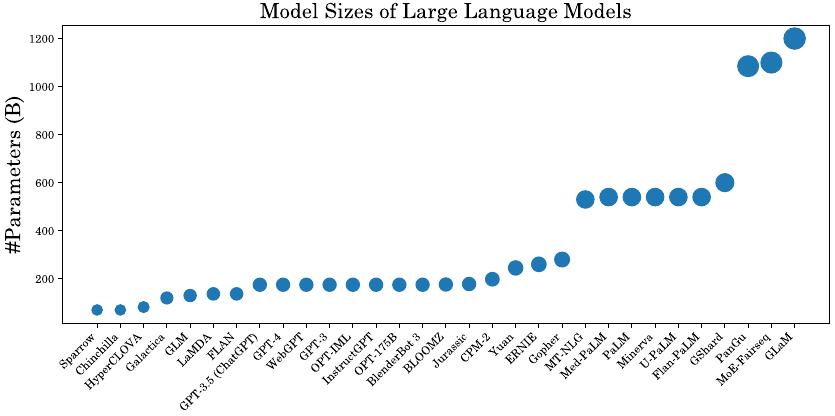}
\caption{Model size of state-of-the-art LLMs \cite{39}}
\label{fig:llms39}
\end{figure}

Recent advances in natural language processing (NLP) have been notably marked by the development of large language models (LLMs) such as the GPT \cite{gpt} class of AI systems. Designed to process, understand, and generate human language with unforeseen complexity, these models have set new computational benchmarks. Trained on extensive datasets covering a wide range of language-related tasks, the rapid evolution of these models has been driven primarily by progress in model architecture, training methodologies, and the increasing availability of large-scale computing resources \cite{7}. \\
As a result, the pace of development and the complexity of these systems have grown at an unprecedented rate. Academic institutions and private AI research groups strive to improve performance across a suite of NLP benchmarks, driven by the principle that significant advancements in deep learning are achieved by scaling larger neural networks with more data and increased computing power. This trend has been conceptualized by \textit{neural scaling laws}, which empirically state that modeling performance improves as model size, dataset size, and computational resources are scaled \cite{scaling}. Consequentially, models containing millions to billions of parameters are trained on huge data center clusters, typically based on some variant of the original transformer architecture proposed in \cite{att}. \cite{43} quantifies this development by the size of transformer-based language models having increased tenfold yearly for the past several years, a trend similarly reflected in Fig. \ref{fig:llms39}. \\
This exponential growth in model size is anticipated to continue as long as system and hardware technology can keep pace. In fact, these models' associated high hardware costs have hindered their wider adoption so far, necessitating more efficient hardware designs in the future. To date, executing these expensive models often involves several thousand GPUs (graphics processing units) and requires large spending, highlighting the significant investment and need for technological advancements in the field \cite{9}.

\subsection{Scaling Challenges in Neural Networks}
Contrary to the remarkable progress in software, hardware development has struggled to keep pace with the computational demands imposed by ever-growing models, particularly in terms of computational and energy costs required for inference \cite{7}. This intensity is primarily due to the scale of data being processed by the transformer architecture, which is central to these models yet comes at the cost of intensive computation and frequent memory accesses. \\
Historically, though, this scaling hypothesis in software has been supported by technological advances in transistor design, which have resulted in faster and more energy-efficient components, providing ever-increasing speed and storage capacity \cite{50}. However, Moore's Law, stating circuit density doubles every two years, has slowed, revealing the limitations of the traditional discrete structure of memory and processing units \cite{biologically}. To date, modern computing systems remain predominantly processor-centric, with data processing confined to CPUs, GPUs, FPGAs, or ASICs. Data must be moved from its origin to these processors, which severely limits performance, energy efficiency, scalability, and ultimately sustainability \cite{65}. Characterized by these inefficiencies in the constant data movement between memory and processing units, this structure has led to the well-known von Neumann bottleneck. Studies indicate that even the most powerful processors and accelerators waste a significant share of time - up to more than 60\% - waiting for data, despite multiple levels of caches, DRAM chips, memory systems, and interconnects \cite{61}. In fact, traditional architectures consume most of their energy in moving data, which is orders of magnitudes larger than the energy used for the actual floating point operations (FLOPs) \cite{66}. With GPUs being notably non-scalable in power and cooling requirements, this energy bottleneck leads to low utilization rates. Thus, GPUs have scaled greatly in capability and quantity but only modestly in energy efficiency \cite{50, 39}. These computational challenges lead to higher latency, complicating real-time and resource-constrained applications and, hence, the broader adoption of these technologies \cite{61}. As systems scale up to accommodate models the size of GPT, this performance gap only widens - thereby driving the need for more energy-efficient hardware solutions. \\
Here, inference tasks, which often receive less attention than training, account for a significant portion of energy costs. This is notable due to LLMs being repeatedly called upon to perform inference as opposed to the one-time resource consumption during training, such as in the case of chatbots like ChatGPT \cite{27}. Estimates from NVIDIA and Amazon suggest that inference tasks amount to 80\% or more of AI compute demand while training new models accounts for a much smaller fraction \cite{77}. Understanding resource utilization during inference is therefore crucial for cost savings, scaling performance, and optimizing hardware deployment. \\
To address the inefficiencies of current system architectures, researchers have explored several techniques to optimize transformer inference at various levels of abstraction. Methods such as pruning \cite{p1, p2}, quantization \cite{q1, q2, q3}, and knowledge distillation \cite{kd} have demonstrated potential to enhance scalability and efficiency by reducing model size and computational requirements. Yet, the physical separation between memory and compute persists, necessitating an orthogonal paradigm shift in processing for more substantial improvements \cite{39}.

\subsection{Efficiency Gains through Compute-in-Memory}
In this light, Compute-in-Memory (CIM) becomes a promising solution that fundamentally reduces the data movement bottleneck. By shortening the data paths and hence minimizing latency, CIM improves both performance and energy efficiency simultaneously. Architectures can substantially accelerate AI inference by performing analog computations directly in memory - a fundamental departure from modern processor-centric systems. Here, multiply-and-accumulate (MAC) operations, which dominate deep learning operations, are optimized by reading the rows of an array of memory cells (SRAM or emerging non-volatile memory (NVM), see section \ref{devices}) and collecting currents along the columns, thus enabling parallel processing of matrix-vector multiplications (MVM). Among these promising technologies, NVM stands out for its high density, fast access, and low leakage, addressing the critical needs for computational density and memory bandwidth essential for LLM efficiency. Given the predominant reliance of transformers on matrix multiplications, they present themselves as excellent candidates for acceleration by advanced CIM hardware. \cite{61, 66}

\subsection{Hardware-Software Co-Design in CIM Systems}
Historically, the miniaturization of transistors has allowed digital hardware, software, and data within computational systems to operate largely independent of physical changes at the device level, whereas CIM now introduces a shift from a digital to a mixed-signal workflow. This requires not only deep expertise in design and specialized fabrication processes but also a re-evaluation of how these systems are interconnected and optimized together. 
In particular, while promising, CIM technology presents challenges such as programming analog values, managing potential inaccuracies over time, and mapping nonlinear operations to the CIM's MAC framework. These issues can complicate system design and impact reliability, leading to the critical question: \textit{How can a software stack effectively bridge the gap between AI models and CIM hardware, leveraging hardware capabilities to optimize large language model acceleration while addressing its prevailing software challenges?} \\
Accordingly, this survey paper synthesizes various experimental approaches targeting different aspects of CIM to target the imminent challenges of today's computing systems and enhance the performance and energy efficiency of LLM inference. By examining numerous accelerator designs and drawing comparisons with traditional GPU implementations, an in-depth analysis of architectures, performance metrics, and energy efficiency considerations shall be provided. This is to offer critical insights for researchers and engineers to enhance the deployment and use of LLMs in real-world applications. \\
The paper is structured as follows: Section II provides an overview of transformer models and the associated computational cores, along with further background on CIM technologies. Section III discusses the open challenges of (transformer-based) CIM architectures, and subsequent sections explore existing efforts in hardware accelerators. The paper concludes with a synthesis of findings and recommendations for future research directions.

%% file: background.tex
\section{Theoretical Background}
\subsection{Transformer Architecture}
\begin{figure}[t]
\centering
\includegraphics[width=0.6\linewidth]{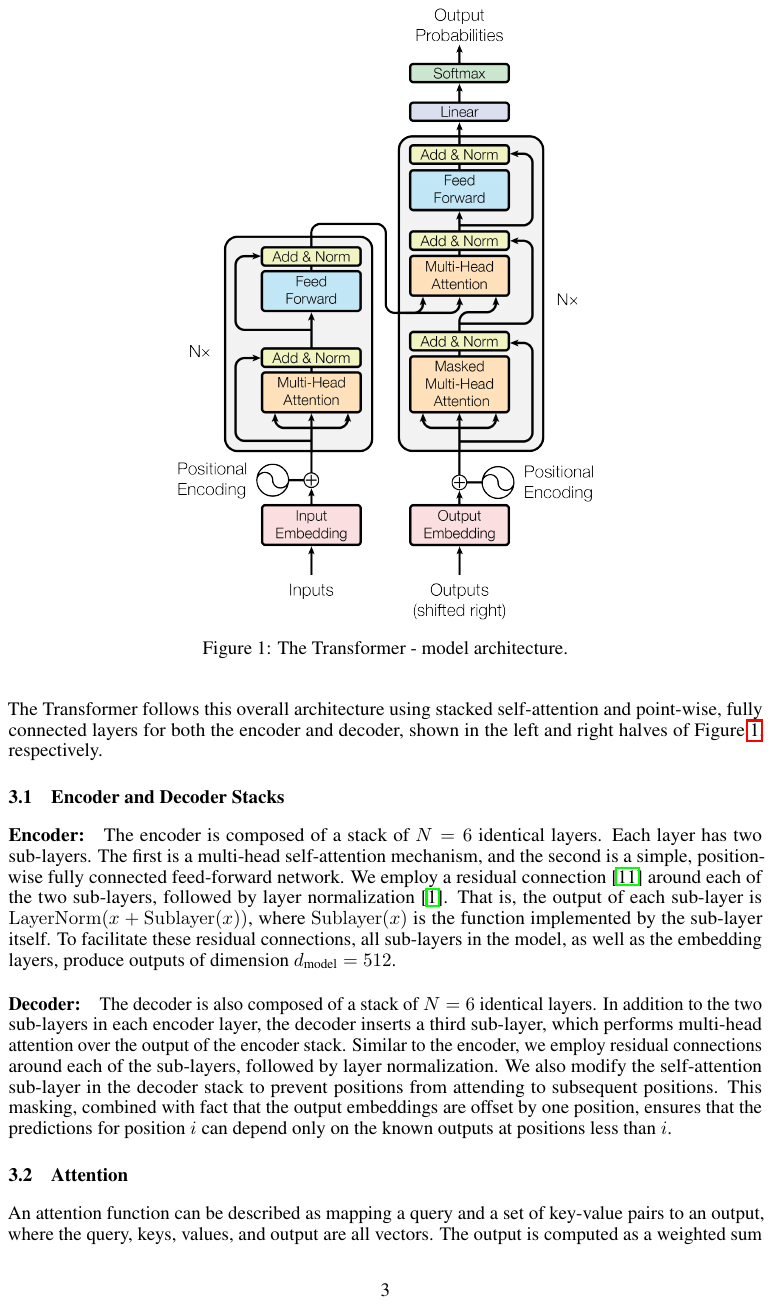}
\caption{The transformer model architecture \cite{att}}
\label{fig:transformer1}
\end{figure}

\begin{figure}[b]
\centering
\includegraphics[width=0.44\linewidth]{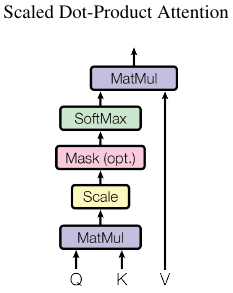}
\includegraphics[width=0.4\linewidth]{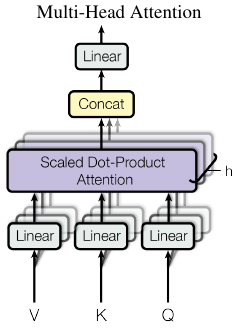}
\caption{Scaled dot-product attention and multi-head attention \cite{att}}
\label{fig:transformer2}
\end{figure}

Modern language models are predominantly built using advanced deep learning techniques, with transformer \cite{att} architectures at their core. They have gained widespread prominence in language processing and computer vision for their ability to process and capture long-range dependencies in input data without the need for recurrent or convolutional layers. Hence, this design allows for highly parallelized computation, optimizing processing speed and scalability. \\
The operational backbone of transformers consists of building blocks such as matrix-matrix projections, attention computations, and feed-forward neural network computations. Therein, they bring key computational advantages such as reduced complexity per layer and the capability to perform operations in a highly parallel manner. In particular, the multi-head attention (MHA) mechanism connects input and output sequences of length $n$ with $O(1)$ operations; a significant efficiency gain over the $O(n)$ operations required by traditional LSTM (long short-term memory) \cite{lstm} or RNN (recurrent neural network) \cite{rnn} architectures for similar tasks. \cite{66} \\
At the hierarchical level, typical transformer models employ an encoder-decoder structure. The encoder consists of a stack of identical layers, each comprising two primary components (see Fig. \ref{fig:transformer1} \& \ref{fig:transformer2}): Said multi-head attention mechanism and a position-wise feed-forward neural network. This configuration allows the model to evaluate the relevance of different parts of the input sequence, capturing essential long-range dependencies. The positional encoding provides information about input elements' relative or absolute positions, compensating for the attention mechanism's lack of inherent positional awareness. Each layer in the encoder further contains a residual connection around each of the sublayers, followed by a normalization layer, which improves stability and performance.
The decoder mirrors this structure but with an additional layer for multi-head attention over the encoder's output, facilitating effective synthesis of the processed information. \cite{att}

\subsection{Attention Computation}
As a key novelty, attention mechanisms in transformers allow the model to focus dynamically on different parts of the input sequence, assigning more computational weight to more relevant parts and significantly improving the model's ability to capture long-range dependencies. \\
Conceptually, the attention mechanism works by computing a weighted sum of a set of input values \(V\) based on their compatibility with a query \(Q\). It computes an attention score for each key-query pair using a scoring function and normalizes these scores using a \textit{softmax} function to produce attention weights. These weights are then used to compute the weighted sum of the values, representing the attention mechanism's output. This mechanism is next extended to a multi-headed approach, detailed in Fig. \ref{fig:transformer2}. Here, the input \(X\) is projected into different subspaces to obtain distinct \(Q\), \(K\), and \(V\) vector representations for each head. Each head independently computes the scaled dot-product attention, allowing the model to simultaneously attend to information from different representation subspaces at different positions. This is achieved by projecting the query \(Q\), key \(K\), and value \(V\) through separate projection matrices \(W_{i}^{Q}\), \(W_{i}^{K}\), \(W_{i}^{V}\), respectively, before concatenating and linearly transforming the results. Ultimately, the model's ability to analyze the relationships between sequence elements in a highly parallelizable manner is enhanced, revealing different aspects of the input through each head. \cite{att, 41} \\
In mathematical terms, each head in the MHA layer, therefore, consists of three trainable matrices: the query matrix $W^Q$, key matrix $W^K$, and value matrix $W^V$, structurally given as linear layers. The corresponding queries $Q$, keys $K$ and values $V$ are computed by multiplying the input sequence $X = [X_0, X_1, ..., X_n]$ with the above matrices as:

\begin{equation}
    Q = X W^Q, \quad K = XW^K, \quad V = XW^V
\end{equation}

It holds $X \in \mathbb{R}^{n \times d_{model}}$, $W^Q, W^K \in \mathbb{R}^{d_{model} \times d_k}$, $W^V \in \mathbb{R}^{d_{model} \times d_v}$, $Q, K \in \mathbb{R}^{n \times d_k}$, $V \in \mathbb{R}^{n \times d_v}$, where $d_{model}, d_k, d_v$ are the dimensions of the model, keys, and values, respectively. \\
Next, the attention score is computed using the scaled dot-product attention layer, i.e.,

\begin{equation}
    \text{attention}(Q, K, V) = \text{softmax}\left(\frac{Q K^T}{\sqrt{d_k}}\right) V
\end{equation}
where the result is $e \in \mathbb{R}^{d_k \times d_v}$. Conceptually, this attention layer is decomposed into matrix multiplications, scaling, and \textit{softmax} functions. \textit{Softmax} for a $d_k$-dimensional vector is calculated as follows:

\begin{equation}
    \text{softmax}(x_i) = \frac{e^{x_i}}{\sum_{j=1}^{d_k} e^{x_j}}
\end{equation}

The final multi-head attention result is obtained by multiplying the concatenation over the attention results of each head and the weight matrix $W^O$, such that

\begin{equation}
    \text{multihead}(Q, K, V) = \text{concat}\left(head_1, ..., head_h\right) W^O
\end{equation}
where $head_i = \text{attention}(Q_i, W_i, V_i)$ and $W^O \in \mathbb{R}^{h d_v \times d_{model}}$. The outputs are eventually passed through the position-wise feed-forward layer according to

\begin{equation}
    \text{FFN}(X) = \max\left(0, X W_1 + b_1\right) W_2 + b_2
\end{equation}
where $W_1 \in \mathbb{R}^{d_{model} \times d_{ff}}$, $W_2 \in \mathbb{R}^{d_{ff} \times d_{model}}$, and the result $\text{FFN}(X) \in \mathbb{R}^{d_{model}}$. $d_{ff}$ is the dimension of the hidden layer.

\subsection{Large Language Models}
This general architecture is exemplified by two highly influential models: BERT \cite{bert} (Bidirectional Encoder Representations from Transformers) and GPT \cite{gpt} (Generative Pre-trained Transformer). BERT employs an encoder-only architecture that processes all input tokens simultaneously, ideal for tasks requiring a comprehensive understanding of contextual relationships. Conversely, GPT uses a decoder-only architecture, processing tokens sequentially to predict subsequent tokens based on previous ones, making it particularly suited to generative tasks. Both models consist of $N$ identical blocks, including an attention module and a feed-forward network, highlighting transformer architectures' modular and scalable nature. 
However, the sequential processing of GPT results in notable underutilization of GPUs, especially for small-batch inference tasks \cite{22}. Compared to CNNs (convolutional neural networks), GPT models are characterized by their substantial sizes and low compute-to-memory ratios. For instance, GPT-3 contains 175 billion parameters with an input sequence of 2048 \cite{gpt}. This means that large amounts of data must be accessed through off-chip memory, resulting in performance and power consumption penalties.

\begin{table}[h]
\centering
\caption{Transformer-Based Large Language Models \cite{76}}
\begin{tabular}{|l|c|c|c|}
\hline
 & \textbf{BERT-base} & \textbf{BERT-large} & \textbf{GPT-3} \\
\hline
\textbf{Type} & Encoder & Encoder & Decoder \\
\textbf{Number of layers} & 12 & 24 & 96 \\
\textbf{Weight count} & 85M & 302M & 174B \\
\textbf{Sequence-length} & 128--512 & 128--512 & 2,048 \\
\textbf{FC : Att. comp.} & 97\% : 3\% & 98\% : 2\% & $>$95\% : $<$5\% \\
\hline
\end{tabular}
\end{table}

As the demand for more complex LLMs increases, there is an urgent need to address the computational challenges associated with their size and complexity. Primarily, these are credited to the significant communication overhead for transferring parameters from memory to compute units, creating a substantial bottleneck. For instance, while multi-head attention can achieve a constant complexity of \(O(1)\), this is contingent upon sufficient parallelization - a factor often limited by the memory bandwidth, with the memory requirements growing quadratically with sequence length \(n\). Yet, the ability to handle longer sequences has become increasingly important with the growth of NLP datasets, with early models having millions of parameters, GPT-3 billions, and Switch Transformers even reaching trillions \cite{38}. Thus, implementing these mechanisms in hardware requires careful adaptation to fully exploit their potential.

\subsection{Conventional Approaches to LLM Acceleration}
This challenge has led to the development of specialized processing units or dedicated hardware units to perform operations more efficiently. Processors like Google's TPUs (tensor processing unit) and AWS Inferentia cater to specific computational needs, while GPUs offer a more general-purpose solution, indicating a diverse and evolving market. Nevertheless, several bottlenecks continue to hinder their full potential. One critical issue is that the unique scaled dot-product attention mechanism, combined with the presence of many fully connected (FC) layers and intensive memory accesses, leads to low arithmetic intensity, measured in FLOPs per byte, which prevents full utilization of the compute units \cite{16, 38, 68}. This, in turn, results in long execution times due to the large memory footprint and low data reuse rate, stressing the memory system while underutilizing the compute resources \cite{32, 62}. Therefore, increasing compute density and memory bandwidth are critical to meeting the growing demands of these complex models for the efficient operation of LLMs. \\
Still, transformers are commonly implemented on GPUs, wherein the computational complexity is limited to \(O(dn^2/c)\), with \(n\) being the sequence length, \(d\) the feature embedding dimension, and \(c\) the number of parallel cores \cite{32}. Therefore, large increases in sequence length and parameters will increase latency, memory bandwidth, and power due to their quadratic time and space complexity. \\
As a consequence, several algorithm-based acceleration techniques have been developed to mitigate these challenges. For example, quantization reduces precision to alleviate memory requirements and latency by minimizing data transfer between compute and memory, with 8-bit precision shown to be an ideal tradeoff while maintaining accuracy \cite{32}. Attention caching reuses attention keys and values computed in previous time steps in autoregressive decoder layers, significantly improving computational efficiency for long sequences \cite{32}. In addition, model parallelism distributes the network across multiple processing units to reduce time complexity and improve utilization, with the Megatron model \cite{mega} achieving 30-52\% of the theoretical peak FLOPs on 512 GPUs \cite{38}. \\
On the other hand, sparse attention mechanisms focus solely on the most relevant keys and values for a given query, reducing computational complexity. Techniques such as locality-based or content-based sparsity, including sliding windows for character modeling or clustering for similarity-based attention, reduce the quadratic complexity of full attention to more manageable levels \cite{32}. Knowledge distillation \cite{kd} trains smaller student models from larger, complex LLMs to maintain performance while reducing computational demands. Meanwhile, compiler optimizations, such as those included in NVIDIA's Cuda or AMD's ROCm, translate code into highly performant versions tailored to specific hardware, enhancing efficiency for LLM workloads. 

To conclude, despite their recent advances, memory and compute bottlenecks prevent transformer networks from scaling to long sequences due to their high execution time and energy consumption \cite{32}.
While software methods such as quantization and pruning help reduce workloads, they can only complement hardware improvements. Therefore, in the spirit of Richard Sutton's \textit{Bitter Lesson} that ``the only thing that matters in the long run is the leveraging of computation" \cite{bitter}, we want to focus on the orthogonal advancements in innovative architectural solutions that can better integrate memory and computational processes to improve the efficiency and scalability of transformer models.

\subsection{Compute-in-Memory}
As computational bottlenecks have been identified, particularly in the self-attention operator and fully connected networks, CIM can alleviate these memory bottlenecks by reducing transfer overhead between memory and compute units, thereby allowing transformers to scale to longer sequences. \\
Conceptually, this computing paradigm performs operations within the memory array and integrates dedicated function units for digital computation to enable end-to-end network inference, hence significantly minimizing the number of off-chip accesses. It offers high memory density and extensively parallel matrix-vector multiplications \cite{biologically}. These benefits are especially pronounced for workloads with large, fully connected layers, such as those found in transformer-based models with layer stacks, each containing four large FC layers (in-projection, out-projection, FC1, FC2) and a self-attention block \cite{76}. While the FC layers typically dominate the workload and scale linearly with sequence length, the attention operations are gaining importance due to their quadratic scaling, as shown in the time distribution in Fig. \ref{fig:mha38}. Large MVMs in these operations are computationally expensive for both conventional and custom digital accelerators but ideally suited for analog NVM \cite{57}. 

\begin{figure}[b]
\centering
\includegraphics[width=0.75\linewidth]{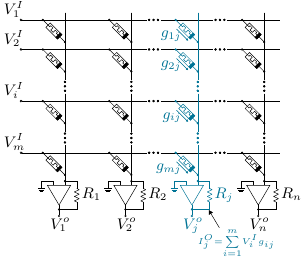}
\caption{Memristor crossbar architecture \cite{crossbar}}
\label{fig:crossbar}
\end{figure}

At their core, NVM arrays are arranged in two dimensions and programmed to discrete conductances (Fig. \ref{fig:crossbar}). Each crosspoint in the array has two terminals connected to a word line and a bit line. Digital inputs are converted to voltages, which then activate the word lines. The multiplication operation is performed between the conductance $g_{ij}$ and the voltage $V_i$ by applying Ohm's law at each cell, while currents $I_j$ accumulate along each column according to Kirchhoff's current law: 

\begin{figure}[t]
\centering
\includegraphics[width=\linewidth]{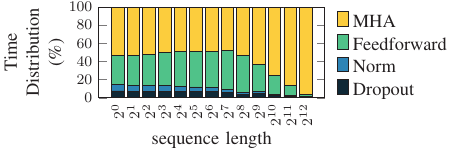}
\caption{Distribution of execution times of the different operations in transformer networks at varying sequence lengths \cite{38}}
\label{fig:mha38}
\end{figure}

\begin{equation}
    I_j = \sum_{i=1}^n V_i g_{ij}
\end{equation}

With this corresponding to a dot product of column conductances and the input voltage, the current is converted via analog-to-digital converters (ADCs). By simultaneously obtaining currents in multiple columns, this process efficiently performs a MVM with just a single read operation \cite{33, biologically}. \\
With computations performed at the location of the data, the von Neumann bottleneck is mitigated since all necessary memory is on-chip, eliminating external data fetch delays. As a result, memory access complexity is reduced from \(O(n^2)\) to \(O(n)\) and MVM computation to constant time \cite{22, 23}. Larger networks are supported by segmenting them into smaller sub-networks across multiple chips. However, CIM arrays will likely need to be sized, interconnected, and optimized for specific inference models to achieve better utilization and efficient weight programming, necessitating further conceptual considerations introduced in section IV. 

\subsection{Memory Device Technologies} \label{devices}
Thus, to effectively implement CIM, a range of technologies has been proposed, including commercial memories such as SRAM, DRAM, and flash memory, as well as emerging NVM like ReRAM (resistive random-access memory), PCM (phase change memory), FeFET (ferroelectric field-effect transistor), and MRAM (magnetoresistive random-access memory). Latter devices typically offer higher performance, density, and lower power consumption, with the added benefit of nonvolatility, which allows continuous operation without the need to permanently power the chip (at the expense of retention) \cite{55, 73}. Yet, compared to CMOS-based approaches, they face challenges such as higher write times and energy requirements, making them less ideal for high-write scenarios \cite{32}. Further, process variations also degrade sensing margins. On the other hand, while more mature due to continuous transistor scaling and hence reduced size and cost, CMOS still suffers from high leakage power and the need for periodic refreshing, making it challenging to apply to large data-centric workloads. \\
A similar distinction can be drawn between digital and analog designs. In digital CIM, one device corresponds to one bit, offering high accuracy but limited throughput due to the small number of rows that can be operated simultaneously without degrading signal quality, limiting weight density. Analog CIM, on the other hand, offers higher weight density and the ability to operate more rows simultaneously. However, it suffers from noisy weights and higher latency due to the need for high-resolution ADCs \cite{55}.
Accordingly, the following details a comparison of these device technologies, with a more quantitative analysis provided in Table II.

\begin{table}[h]
\centering
\caption{Comparison of CIM Technologies based on \cite{71, 32}}
\begin{tabular}{|l|c|c|c|c|c|}
\hline
 & \textbf{SRAM} & \textbf{ReRAM} & \textbf{PCM} & \textbf{FeFET} & \textbf{MRAM} \\
\hline
\textbf{Multi-level} & No & Yes & Yes & Yes & No \\
\textbf{Cell area} & 160F\textsuperscript{2} & 16F\textsuperscript{2} & 16F\textsuperscript{2} & 16F\textsuperscript{2} & 30-80F\textsuperscript{2} \\
\textbf{Weight density} & Low & High & High & High & High \\
\textbf{R\textsubscript{ON}/R\textsubscript{OFF}} & Low & High & High & High & Low \\
\textbf{Write energy} & $<$0.1 nJ & 2 nJ & 6 nJ & 0.1 J & $<$1 nJ \\
\textbf{Write latency} & $<$1 ns & 100 ns & 150 ns & 175 ns & $<$1 ns \\
\textbf{Leakage power} & High & Low & Low & Low & Low \\
\textbf{Endurance} & $>$10\textsuperscript{16} & 10\textsuperscript{5} & 10\textsuperscript{7} & 10\textsuperscript{10} & 10\textsuperscript{15} \\
\hline
\end{tabular}
\end{table}

\subsubsection{CMOS} 
The conventional charge-based CMOS approach uses 6T to 12T SRAM-type structures, which are easier to fabricate but larger. This is why, even at advanced technology nodes, off-chip weight buffers are needed to hold network weights, limiting efficiency gains to small networks that can fit entirely on the chip. To tackle this, first hybrid designs also exploit DRAM's high storage capacity, allowing all model parameters to be stored on-chip. Still, these designs face scaling challenges and thermal noise issues, with capacitor size variations limiting the achievable precision. \cite{65}
\subsubsection{ReRAM} In contrast, NVM technologies store network weights directly on-chip due to their high storage density. Here, ReRAM is among the most promising NVMs because of its low access energy, multi-level cell capabilities, and 3D integration. The 1T1R (one transistor, one resistor) structure of resistive RAM is small, and extensive research is underway to optimize its fabrication and control even further. \cite{57} 
\subsubsection{PCM}
Similarly affected by limited write speed, PCM leverages thermally driven reversible transitions between amorphous and crystalline states, resulting in variable conductance states \cite{57}. Widely researched (\hspace{1sp}\cite{55,57,65,76}), PCM devices are characterized by high on-resistance and high on/off ratios, which facilitate the storage of intermediate resistances even at an array level, despite having slower access times and lower endurance compared to MRAM \cite{65}. This, combined with its ease of fabrication, makes PCM a compelling choice for memory technologies.
\subsubsection{FeFET} Unlike the before-mentioned, FeFET does not require high write currents for programming. With its three-terminal transistor, it structurally resembles a MOSFET and offers better cycle-to-cycle variation and higher endurance of up to possibly \(10^{12}\) according to experimental data \cite{73}. The integration of FeFET into CMOS fabrication processes and its high read density make it suitable for memory-intensive applications such as transformers \cite{32}.
\subsubsection{MRAM}
Lastly, MRAM utilizes magnetic states to store data, offering high endurance compared to other device technologies. Though less widespread, MRAM is notable for its robust cycling endurance while facing on/off ratio limitations and a lack of support for multi-level representations. Additionally, MRAM requires a larger area per bit, which presents challenges for scaling in dense memory arrays. \cite{65}

In conclusion, CMOS offers low write latency and power, making it ideal for transformer sublayers with frequent write operations. However, its leakage power is detrimental for storing static weights. On the other hand, NVM technologies store static weights without the need for periodic refreshes and offer high density, making them apt for applications requiring large amounts of memory and efficient computation \cite{55}.

%% file: challenges.tex
\section{Design and Reliability Challenges}
Based on this diverse stack of technologies, various chips with fully integrated data conversions have been fabricated, achieving competitive energy efficiencies of more than 10 $TOPs/W$ for reduced-precision MVM. Recently, also multicore chips supporting larger networks have demonstrated promising inference accuracy and high energy efficiency. \cite{55} \\
Yet, while the macro-level energy efficiency of CIM tiles is demonstrably high, achieving equally high system-level results is non-trivial and requires careful design considerations. Factors such as signal path crosstalk, leakage current from adjacent devices, design choices like the number of rows turned on simultaneously, ADC precision, crossbar array size, multi-bit storage characteristics, area, write energy, and latency all play a critical role \cite{23}. Reliability issues and the area and energy overhead of ADCs must also be addressed. Other challenges include write persistence/error rate, thermal stability, reliability, endurance, and resistance drift. \cite{71} \\
These challenges and their impact on the design and operation of CIM systems are discussed in the following sections.

\subsection{Analog Computation}
Due to their analog nature, NVM crossbars inherently possess non-idealities that lead to imprecision and erroneous computations. Key issues include read noise, programming errors, conductance drift, and system noise due to MVM circuit integrations such as ADC quantization and dynamically computed IR drops. Here, programming noise is the error introduced when encoding weights in the device; instead of achieving the correct target, the final conductance has an error model based on the standard deviation of iteratively programmed conductances via a normal distribution. Resistance drift, on the other hand, causes conductance states to decay toward zero. Instabilities, such as read noise, lead to slightly different evaluations, even in the absence of programming errors or drift. These errors originate from source and sink resistances in peripheral circuits, parasitic wire resistance in metal traces in crossbars, and sneak paths in circuits. \cite{57}
Further challenges are posed by hard faults, with mitigations proposed to tolerate or write-verify the effects of hardware failure, often resulting in memory overhead and additional NVM/ADC modules \cite{67}. These, in turn, increase write time and reduce endurance, creating challenges for high-write scenarios. \\
Beyond this, the on/off ratio affects computation and quantization errors in the output, with a high ratio being desired to distinguish between the on and off states and to reduce the effects of non-idealities on the output. \cite{71} \\
With process variations and write nonlinearity, careful consideration of inaccuracies is crucial in CIM designs. This is seen as the cumulative effect of both static and dynamic errors can result in a significant loss of accuracy. In lower device variation regimes (0-0.2), there is less than 1\% accuracy loss. However, larger variations can result in up to 4.2\% accuracy loss for BERT-base and 9.87\% for BERT-large due to the accumulation of errors over many layers. \cite{33} \\
Thus, alongside materials research on devices with lower process variation, techniques like remapping, threshold training, and innovative redundancy strategies have been proposed to mitigate device variation and crossbar noise, further explained in later chapters. \cite{23, 60, 67}

\subsection{Peripheral Overhead}
Inherently tied to the analog computation scheme, the complexity of the associated ADCs is a significant challenge in CIM architectures. In ReRAM-based designs, ADCs account for up to 30\% of area and 50\% of the power as measured in \cite{78}. \cite{32} reports figures as high as 58\% (energy) and 81\% (area), with the energy increasing exponentially with higher precision. 
Solutions for fully analog peripherals exist but come at the cost of flexibility and accuracy. Multiplexing ADCs increases latency, and the finite resistance of crossbar wires causes parasitic voltage drops during readout when high currents flow through them, limiting the maximum crossbar size that can be reliably operated and integration density. \cite{65} \\
It is, therefore, vital to reduce the peripheral overhead to make CIM more efficient. One approach is to reduce the ADC precision while maintaining computational accuracy. Here, software co-optimization to reduce ADC precision can be formulated as a quantization problem, where partial sums are also considered in addition to commonly quantized data such as weights and activations. Several training methods have already been proposed to achieve high accuracy with reduced ADC precision. \cite{71}

\subsection{Limited Precision}
Closely related but with implications in the opposite direction, the analog nature of computation and bit precision in CIM pose additional challenges. In particular, it remains difficult to provide accurate floating-point operations because of the limited resolution of multilevel NVM cells in MVMs \cite{66}. However, AI applications have been demonstrated to operate at continuously lower precision: Quantitatively, requirements have dropped from 32-bit to 16-bit, 8-bit, and even down to 1-bit to represent the parameters of neural networks. Weight quantization to as few as 2-4 bits can often be tolerated without significant loss of accuracy, although more training cycles may be required for reliable inference at low precision \cite{81}. Recent developments even introduced quantization schemes tailored specifically for LLMs, further reducing precision requirements, as demonstrated in \cite{qlora}.
At the same time, ReRAM has demonstrated programmability of up to 8 bits but often falls short in large arrays due to stochastic variations, such as drift and noise \cite{78}. Related to the peripheral problems, bit slicing and partial sums, stored in two cells located in adjacent columns, would require shifting and adding modules to achieve higher precision indirectly. 

\subsection{Endurance}
\begin{figure}[t]
\centering
\includegraphics[width=0.95\linewidth]{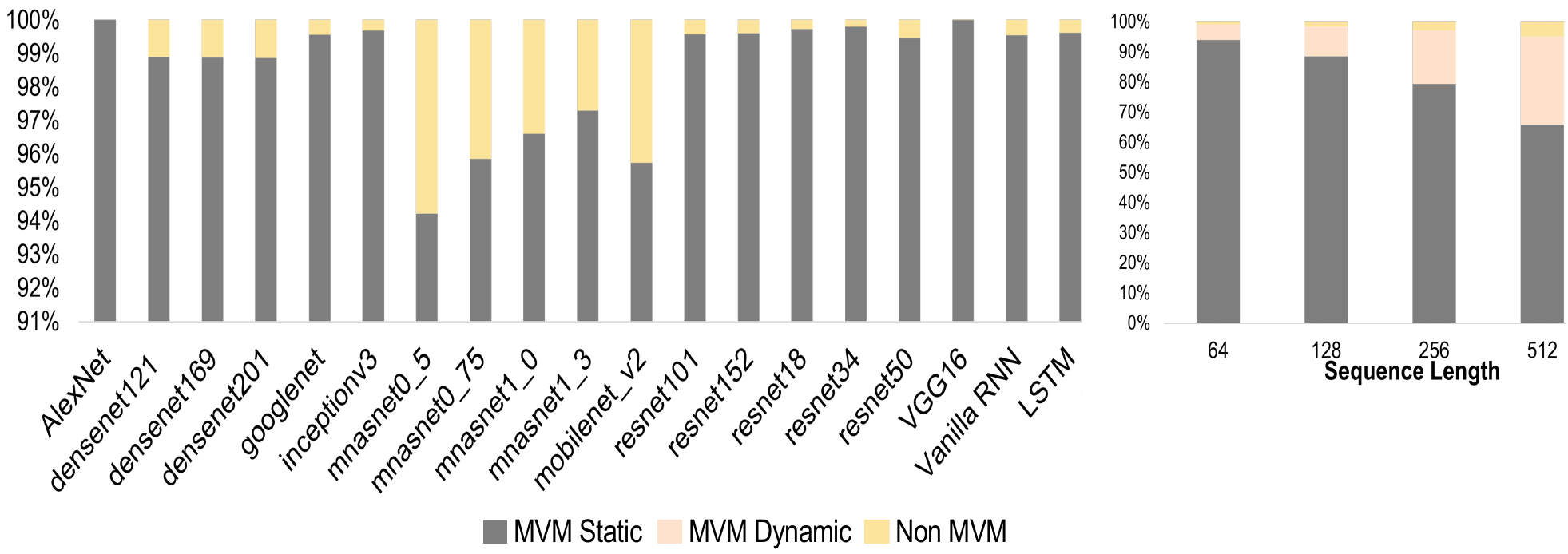}
\caption{FLOP distribution trends for conventional image classification networks (left) compared to transformers (right) \cite{33}}
\label{fig:static33}
\end{figure}

Another challenge unique to CIM relates to finite endurance and the slow, power-hungry nature of weight programming. Fully weight-stationary approaches, where weights are pre-programmed before the inference workload is executed, have been shown to be up to 140x more energy efficient than NVIDIA A100 GPUs \cite{76}. However, data-dependent/dynamic MVM operations, where an operand needs to be written to the crossbar first, result in long write latency and low endurance. It is there that the computational kernels for transformers differ from traditional image classification and previously proposed language models: In classical models, MVMs between inputs/activations and static weights are not reprogrammed, making static MVM operations comprise 95\% of total FLOPs. \\
In contrast, transformers also require MVMs with dynamic query, key, and value vectors that change with each input, reaching only 65\% for a sequence length of 512, with a decreasing trend as shown in Fig. \ref{fig:static33}. Thus, dynamic operations are becoming increasingly dominant and represent a bottleneck in energy and lifetime for crossbar architectures. \cite{33} Despite efforts to improve device lifetime, endurance remains an open problem in modern in-memory design. However, rapidly advancing fabrication and compilation techniques promise to significantly enhance reliability \cite{23}.

\subsection{Computation}
While researchers have already applied CIM to accelerate CNNs and RNNs due to their structural regularity, the unique computational process of attention poses additional challenges \cite{41, 63}. Transformers involve more complex operations such as attention mechanisms, layer normalization, multi-head attention, and softmax operations, making the data flow more complicated than CNN's \cite{39}. Exemplary, the variable sequence length of each input complicates the design because the size of crossbars cannot be fixed, so more rows and crossbar arrays are needed to provide a partial sum that must be aggregated \cite{32}. Also, as introduced earlier, attention score computations require MVMs in which both operands change dynamically for each input, leading to write latency and energy problems for most NVM technologies \cite{33}. Conversely, when multiple input features require identical weight parameters for operations, the CIM-based accelerator accesses the identical arrays serially, producing output sequentially over different time steps and each transformer block becoming a pipeline stage. Therefore, the challenge is that in a self-attending layer and the previous layer, the former remains stalled until the previous layer's computation is complete, causing pipeline bubbles \cite{68}. This is why, to maintain speedups and energy-efficient execution, CIM requires efficient task scheduling approaches \cite{74}. \\
Besides, the crossbar structure becomes unsuitable for nonlinear computations, which constitute up to 27\% of total operations in models like GPT-3 \cite{79}. Complex operations such as reduction and softmax to convert similarity into attention scores require both parallel computation and fine-grained intra-memory data reorganization, which can become performance bottlenecks \cite{31, 68}. Although not a significant concern in digital systems, nonlinear operations can consume up to 70\% of the total system energy in ReRAM, involving ADCs, DACs, and integrated circuits. \cite{78, 79} Lastly, it has been demonstrated that CIM accelerators struggle to support irregularly distributed and varying sparsity patterns, commonly used as a software approach to reduce computational load \cite{70}.

In conclusion, challenges such as non-idealities coupled with the early state of chip fabrication need to be highlighted. Given the inherent nonideal characteristics and their impact on device performance, addressing these design challenges is crucial for advancing the deployment of CIM technologies in AI applications. Novel approaches to mitigate these challenges are to be presented in the following section.

%% file: approaches.tex
\section{Overview of Strategies in CIM Acceleration}
\subsection{Algorithmic Enhancements}
Proposed works in CIM software co-design primarily focus on improving the overall utilization by implementing token pruning to discard less significant tokens and integrating model-adaptivity to reduce idle time. Concepts, such as bitwidth balancing, are integrated into architectures to enhance performance, as seen in \cite{42}. Therein, MulTCIM introduces long reuse elimination by dynamically reshaping the attention pattern to improve CIM utilization. A token pruner removes insignificant tokens to further improve efficiency. The CIM architecture is integrated bitwise as digital MACs in SRAM, eliminating the need for ADC and avoiding nonideality issues caused by signal margins and process variations. However, previous discussions on the advantages and disadvantages of digital CIM apply.
ReBERT \cite{68} builds on similar concepts by addressing the challenge that low attention tokens have relatively little impact on attention results. Thus, excluding these words from self-attention calculation does not degrade accuracy. It utilizes window self-attention, where self-attention computations are focused on words near the target word, reducing the attention scope and, thus, computation time. \\
Introduced in \cite{62}, the TransPIM architecture targets comparable software-side improvements. Using token-based dataflow reduces data movement between layers and implements lightweight hardware modifications in high-bandwidth memory (HBM) to improve computational efficiency and data communication. The architecture describes, therefore, rather a near-memory system with an optimized data communication architecture, where HBM attempts to overcome the time and distance barriers to memory access, but the physical separation between memory and compute units persists \cite{65}. Nevertheless, interesting algorithmic improvements should be considered for future CIM approaches. In particular, efficient dataflow mapping computations to the memory-based architecture use a token-based sharding mechanism, dividing input tokens into different shards and allocating them to different memory partitions. This approach avoids memory traffic for reused data, reduces data movement costs, and exploits more parallelism because different banks can handle computation and data movement for allocated tokens independently. \\
ReTransformer \cite{41}, on the other hand, proposes a ReRAM-based acceleration by using matrix decomposition in the attention mechanism, eliminating data dependencies and reducing computational latency present. The design uses a submatrix pipeline to further improve data throughput. By avoiding intermediate result storage, a common problem in existing CIM, rewrites are reduced and, hence, energy and endurance issues, allowing full implementation in ReRAM. \\
Lastly, focusing on the hardware shortcomings of existing designs, \cite{60} introduces ATT as a fault-tolerant ReRAM accelerator designed for attention-based neural networks. It incorporates a dedicated pipeline design and fault tolerance to address hardware inefficiencies by introducing non-uniform redundancy to target hard faults. 

\subsection{Resilience and Fault Tolerance}
Building on this, a combination of differentiable structure pruning, weight duplication, and MSB (most significant bit) embeddings is introduced in \cite{67}. The contribution of this strategy is a fault-tolerant approach without adding additional parameter space. It includes an automatic structured pruning method to reduce parameter space and a weight voting strategy to improve network robustness. Structured pruning reduces model redundancy by dropping entire weights in rows or columns of the weight matrix, creating a row- or column-sparse model architecture. The freed space can duplicate the MSB of important weights, enhancing resilience.
This is to address hard faults, which result in a permanently fixed high or low resistance state due to fabrication defects and immaturity, and amplify in impact as errors accumulate and propagate during forward propagation. This approach has been tested on several tasks from the GLUE benchmark \cite{glue} using the BERT model, demonstrating effective fault tolerance. \cite{67} 

\subsection{Hardware-Aware Training}
A software-side attempt to mitigate these hardware nonidealities is summarized by the promising approach of hardware-aware training. Networks are trained hardware-awarely by injecting noise into the synaptic weights to improve their resilience to nonidealities before deployment \cite{55}. Studies by \cite{81} and \cite{76} demonstrate the robust deployment of various AI workloads on analog in-memory computing by evaluating accuracy impacts and system robustness across multiple network topologies. 
Here, \cite{57} suggests that large neural networks can achieve iso-accuracy with floating-point (FP) implementations (within 1\% of the reference) when properly retrained to account for CIM nonidealities, even after conductance drifts over time. Training is initially done in 32-bit FP precision, followed by retraining with noise (nonidealities, programming and read noise, drift) introduced in the forward pass. In contrast, the backward pass remains in full precision. Drift compensation is achieved using a global correction factor calculated from the mean drift over time and applied by adjusting the digital output scale factor by the delayed and initial inference values ratio. This process is performed once and is not specific to device or chip characteristics, resulting in broadly applicable models. As a result, it enhances signals without removing underlying noise sources yet maintaining competitive accuracy.

\subsection{High-Precision Techniques}
Therewith connected, high-precision floating-point computation in NVM remains challenging due to the resolution limitations of multilevel cells. RIME \cite{66} addresses this by utilizing single-cycle NOR, NAND, and minority logic for in-memory floating-point operations by simplifying peripheral circuitry. Within the ReRAM-based CIM architecture, a central controller manages the switching between different function modes, preprocessing and transferring data, and communicating with external units. CMOS computing units handle calculations of exponential functions and divisions, as they are too complex for the crossbar architecture. Tiles are controlled by a CIM controller to perform parallel multiplication and facilitate 32-bit operations, enabling a full transformer accelerator that achieves 2.3x faster speed and 1.7x more energy efficiency without loss of accuracy.

\subsection{Comprehensive Full-Circuit Design}
With peripheral overhead - strongly influenced by ADC and hence weight precision - several works have attempted to mitigate the dependence on power- and area-hungry ADCs. \cite{75} introduces a shared ramp and compact dedicated comparators to convert analog voltages instead of traditional ADCs. Meanwhile, \cite{23} brings up the idea of bit-level sparsity through bit-mapping and flipping schemes to reduce the need for peripheral circuitry. \cite{82}, finally, takes this a step further by proposing a full-circuit design leveraging memristor crossbar modules and an analog signal memory module for function circuits such as \textit{softmax} and \textit{relu}, as well as a timing signal generator, all of which work without the need for analog-to-digital converters. The four modules (memristor crossbar, analog signal memory, function circuit, timing signal generator) work together to handle weight representation, calculation, intermediate result storage, signal transformation, and operation scheduling. However, this approach has been limited to character recognition and MNIST \cite{mnist} tasks. \\
Similarly, \cite{79} presents an approach using ReRAM for both matrix multiplication and activation functions, extending the versatility of ReRAM toward nonlinear operations. The design provides a comprehensive simulation framework, that reports superior computational efficiency compared to both analog and digital programmable solutions. \\
RACE-IT \cite{78} instead uses analog content addressable memory (CAM) for efficient computation of transformer operations. The efficient execution of all operations within transformer models replaces ADCs by processing analog inputs and producing digital outputs, thus overcoming the difficulties of area and energy-intensive ADCs. 
\cite{70} achieves similar prospects through its time-domain winner-take-all circuit that eliminates ADC requirements. It amplifies the time-delay differences generated by the ReRAM crossbar output and searches for a winner via a binary tree topology. It also splits the architecture into analog ReRAM for token relevance prediction and digital SRAM for accurate self-attention computation due to its precision and reliability.

\subsection{Heterogeneous Computing}
Specifically, this pattern of assigning workloads to different types of CIM devices is called hybrid or heterogeneous architecture. This goes back to transformers exhibiting significant heterogeneity in their compute cores, requiring the integration of different types of hardware modules into a single system. The corresponding computing systems are designed for algorithmic completeness and obtaining configurable precision techniques to manage the precision-efficiency trade-offs of SRAM and NVMs \cite{80}. \\
In \cite{66}, CMOS computing units handle calculations of exponential functions and division calculations, which are too complex for ReRAM crossbars. While most of the exponential operational power is consumed during data transfer, it remains considerably low without significantly impacting overall efficiency. A comparable heterogeneous design achieves 6x the raw throughput and 7x shorter latency, as reported by \cite{76}. An analog ReRAM and digital SRAM accelerator (HARDSEA \cite{70}) achieves even higher numbers using a hybrid computing architecture combined with a product quantization scheme. \\
X-Former \cite{33} further uses a sequence-blocking dataflow to overlap computations across different processing elements. This way, based on CMOS processing tiles and multiple attention computation tiles, the attention engine overcomes low utilization, while NVM tiles handle another projection engine. Therefore, X-Former avoids frequent reprogramming by mapping static operations to NVM while processing dynamic operations in CMOS elements, thus addressing the issues of high write energy/latency and low endurance in NVMs. \\
The development of a hybrid CMOS-FeFET in-memory computing architecture called iMTransformer \cite{32} combines NVM crossbars with CAM instead. It is supported by software improvements through reusable parameters to reduce the number of operations and exploit available parallelism in attention computation. A configurable attention selector chooses sparse attention patterns for energy reduction, and CAM locality-sensitive hashing filters sequence elements by their importance. FeFET crossbars store projection weights, while CMOS crossbars store attention scores for later reuse. Thus, the hybrid setup of projection and attention engines combines the benefits of both NVM and CMOS and significantly improves speed and energy efficiency, as detailed in Table III. \\
Lastly, \cite{43} even introduced a highly heterogeneous GPU extension that uses NVM memory along with GPU to support larger models that exceed GPU memory limits, achieving 25x larger models capable of inference while maintaining over 50\% of peak hardware performance. While not strictly following the lines of compute-in-memory, this approach opens access to large transformer inference by enabling it on systems with limited GPU resources.

\subsection{Innovations in Analog AI Chips}
Combining advances in several techniques described, in recent advancements, IBM has introduced a prototype analog AI chip called NorthPole \cite{54} that achieves comparable accuracy to digital devices while offering faster processing and lower power consumption. Featuring 35 million PCM devices to store neural networks with up to 17 million parameters directly on the chip, NorthPole enables continuous operation without the need for constant power, significantly improving energy efficiency. While not yet comparable to the size of today's most advanced generative AI models, combining multiple chips has enabled real-world use cases to be effectively addressed, with performance rivaling that of digital chips: According to MLPerf benchmark data, the IBM prototype was 7x faster than the best MLPerf submission in the same network category, while maintaining high accuracy and demonstrating approximately 14x better performance per watt. \cite{54}

\begin{table*}[h]
\centering
\caption{Comparison of various CIM architectures for LLM benchmarks}
\begin{tabular}{|l|c|c|c|c|c|}
\hline
 & \textbf{Speedup (Rel.)} & \textbf{Efficiency (Rel.)} & \textbf{Architecture} & \textbf{Benchmark} & \textbf{Model} \\ \hline
\textbf{PIM-GPT} \cite{22} & 41-137 & 123-383 & DRAM & NVIDIA T4 & GPT \\
\textbf{TRANSPIM} \cite{62} & 22.1-114.9 & 138.1-666.6 & DRAM & NVIDIA RTX 2080Ti & RoBERTa, Pegasus, GPT-2 \\ 
\textbf{MulTCIM} \cite{42} & 2.50-5.91 & - & CMOS & SOTA Digital CIM Accelerators & BERT \\ 
\textbf{IMCAT} \cite{38} & 200 & 41 & CMOS & NVIDIA Titan RTX & BERT \\ 
\textbf{PNRU} \cite{79} & 1.51 & - & ReRAM & NVIDIA RTX 3080 & GPT-3 \\ 
\textbf{RIME} \cite{66} & 2.3 & 1.7 & ReRAM & NVIDIA RTX 3080 & Vanilla Transformer \\ 
\textbf{ATT} \cite{60} & 202 & 11 & ReRAM & NVIDIA GTX 1080 Ti & BERT, XLNet, XLM \\ 
\textbf{ReBERT} \cite{68} & 39.2 & 643.2 & ReRAM & NVIDIA Titan Xp & BERT \\ 
\textbf{ReTransformer} \cite{41} & 23.21 & 1086 & ReRAM & NVIDIA Titan RTX & Vanilla Transformer \\ 
\textbf{iMTransformer} \cite{32} & 11 & 7.92 & CMOS + FeFET & NVIDIA Titan RTX & BERT \\ 
\textbf{X-Former} \cite{33} & 85 & 7.5 & CMOS + ReRAM & NVIDIA GeForce GTX 1060 & BERT \\ 
\textbf{RACE-IT} \cite{78} & 10.7 & 1193 & CMOS + ReRAM & NVIDIA H100 & BERT \\ 
\textbf{HARDSEA} \cite{70} & 13.5-28.5 & 291.6-1894.3 & CMOS + ReRAM & NVIDIA RTX 3090 & BERT + GPT-2 \\ \hline
\end{tabular}
\label{table:cim_llm_benchmarks}
\end{table*}

%% file: discussion.tex
\section{Discussion}
Despite the significant progress outlined in developing and deploying CIM technologies, several critical challenges must be addressed to fully realize their immense potential. \\
One major challenge is constituted by the difficulty of scaling up emerging technologies for in-memory computing. For instance, while 224 megabytes of RAM as mentioned in \cite{52} may be sufficient for certain specific applications, it falls far short for large-scale models like GPT, which require several thousand megabytes even in their most scaled-down versions. Hence, scaling solutions to increasingly large transformer models and multi-chip systems without compromising performance and efficiency remains a significant challenge. \\
In particular, most current CIM research focuses on macro-level circuit design. However, as AI models become larger and more complex, optimizing both inter- and intra-CIM architectures is essential to achieve high efficiency at the accelerator level. Future work needs to focus on how to scale out for emerging LLMs and other advanced AI models. \\
Another critical issue is managing the trade-offs between the lower precision of analog computation and the necessity for high precision in certain transformer computations \cite{70}. In this context, full circuit implementations pose several challenges for neural networks due to the large number of MVM operations, intermediate value storage, various function modules, and complex operation scheduling \cite{82}. Further, these implementations lack versatility compared to digital solutions. Designing specialized circuits for each nonlinear function achieves the best energy efficiency but compromises device compatibility to support custom neural network models. Floating-point designs offer versatility at the cost of higher power dissipation due to undesired extra costs such as reconfiguration, routing, and communication \cite{79}. \\
Another less apparent but equally critical challenge is the absence of unified EDA (electronic design automation) tools that can effectively support NVM-based AI accelerators \cite{23}. Such tools are necessary for the exploration, optimization, and deployment of these technologies. While models that facilitate the development of new algorithms and their comparison by establishing a reproducible benchmark are valuable, they cannot replace the ultimate CIM hardware verification of the algorithms \cite{81}. \\
A final challenge is less technical in nature: The widespread adoption of memory-centric computing depends heavily on the need for a better demonstration of its benefits on various workloads, ease of programming, system and security support, and comprehensive infrastructure benchmarking tools. Overcoming the ingrained processor-centric mindset is critical to the future success of memory-centric systems. Thus, further research is encouraged to optimize the trade-off between programmability and efficiency in memristor-based computing systems, ensuring they meet the diverse needs of modern computational tasks.

%% file: conclusion.tex
\section{Future Work}
\subsubsection{Manufacturing Advances} As manufacturing processes for NVM are not yet fully matured, there are significant opportunities to refine these techniques to support more extensive adoption and implementation. Substantial improvements in weight density could be achieved by integrating memory devices closer to the transistor level or by three-dimensional stacking of memory layers, as postulated in \cite{55}. These enhancements could significantly increase the storage capacity, memory access speed, and overall performance of the systems.
\subsubsection{Error Correction} Given the inherent nonidealities that can only be partially overcome, error-correction schemes and fault-tolerance methods are not just beneficial but essential. They play a crucial role in ensuring reliable results for transformer models, thereby highlighting their critical importance in hardware-software co-design for emerging AI applications. 
\subsubsection{Software and Runtime Development} Future work should also focus on the design of sophisticated runtimes and advanced compilation systems that decide which code to execute in CIM units. Developing comprehensive infrastructures and detailed benchmarks will help hardware designers and software engineers accurately assess the benefits, tradeoffs, and overall feasibility of new designs. This focus is essential for streamlining development and enhancing the integration of CIM technologies.
\subsubsection{Co-Design Improvements} Eventually, it is critical to continually refine both hardware and software together to maximize performance and efficiency. This includes exploring advanced model compression techniques coupled with redundancy programming strategies. Designing efficient “software-aware” hardware and “hardware-aware” software will be essential for optimizing the performance and capabilities of AI systems. 
In conclusion, the outlined future work highlights the broad opportunities for further advances in hardware-software co-design. These efforts are critical to realizing the full potential of AI applications, particularly in edge and real-time systems, and could greatly benefit from the ongoing evolution in the field toward the democratization and sustainability of artificial intelligence.

\section{Conclusion}
As we advance in integrating CIM technologies into LLMs, challenges must be addressed to fully realize their potential. Inherent imprecision and the need for new fabrication processes are substantial barriers, but ongoing research is dedicated to overcoming them and enhancing the viability of these technologies. 
Meanwhile, the benefits of advancing CIM technologies are clear, especially as LLMs continue to grow in complexity and scale. These models require vast computational resources to perform inference with minimal latency, thereby enabling increasingly sophisticated capabilities for understanding language with remarkable accuracy. CIM bears the promise to cater to these needs.
In addition, the efficiency of these devices means that they do not require extensive cooling systems to operate, further enhancing their practicality and sustainability. Ongoing advances in CIM technologies promise to revolutionize the deployment of large-scale language models by addressing both performance and environmental concerns. In particular, this survey paper has focused on the synergistic effects of software and hardware co-design to optimize precisely these metrics. We have discussed methods that more efficiently adapt transformers for specific tasks by leveraging the inherent capabilities of CIM hardware. The examples presented validate the effectiveness of a tightly integrated software-hardware approach and show how such integration can optimize transformer models to meet the stringent requirements of modern AI applications. \\
We conclude this paper by emphasizing the future directions in this evolving field of research. Our efforts are intended to stimulate further interest in developing novel optimization techniques that facilitate efficient inference on large language models on CIM hardware, which we see as promising candidates for future computing systems. The development of optimization techniques for efficient transformer inference is now a critical area of research due to its broad applications across different architectures and domains. \\
By bringing together advances in device technology, architectural innovations, and software-hardware co-optimization, we can overcome the current limitations and initiate a new generation of AI systems that are both powerful and efficient.